\documentclass[aps,prl,reprint,superscriptaddress,longbibliography]{revtex4-2}
\pdfoutput=1
\usepackage[colorlinks=true,urlcolor=blue]{hyperref}
\usepackage[normalem]{ulem}
\usepackage{mathtools}
\usepackage{amsfonts}
\usepackage{graphicx}
\usepackage{stmaryrd}
\usepackage{amsmath}
\usepackage{amssymb}
\usepackage{xcolor}
\usepackage{color}
\usepackage{bbold}
\usepackage{bm}

\definecolor{red}{rgb}{0.75,0,0}
\definecolor{blue}{rgb}{0,0,0.75}
\definecolor{green}{rgb}{0,0.5,0}

\newcommand{\blue}[1]{{\color{black} #1}}

\begin{document}
\title{Reply to comment on ``Faceting and flattening of emulsion droplets: a mechanical model''}

\author{Ireth Garc\'ia-Aguilar}
\affiliation{Instituut-Lorentz, Universiteit Leiden, P.O. Box 9506, 2300 RA Leiden, Netherlands}
\author{Ayelet Atkins}
\affiliation{Bar-Ilan Institute of Nanotechnology \& Advanced Materials, Bar-Ilan University, Ramat-Gan 5290002, Israel}
\author{Piermarco Fonda}
\affiliation{Instituut-Lorentz, Universiteit Leiden, P.O. Box 9506, 2300 RA Leiden, Netherlands}
\affiliation{Theory \& Bio-Systems, Max Planck Institute of Colloids and Interfaces, Am Mühlenberg 1, 14476 Potsdam, Germany}
\author{Eli Sloutskin}
\affiliation{Physics Department, Bar-Ilan University, Ramat-Gan 5290002, Israel}
\affiliation{Bar-Ilan Institute of Nanotechnology \& Advanced Materials, Bar-Ilan University, Ramat-Gan 5290002, Israel}
\author{Luca Giomi}
\email{giomi@lorentz.leidenuniv.nl}
\affiliation{Instituut-Lorentz, Universiteit Leiden, P.O. Box 9506, 2300 RA Leiden, Netherlands}

\date{\today}

\newcommand{\lsim}{\raisebox{-0.13cm}{~\shortstack{$<$ \\[-0.07cm]
      $\sim$}}~}

\maketitle

In their Comment \cite{Haas:2021}, Haas {\em et al}. advance two hypotheses on the nature of the shape transformations observed in surfactant-stabilized emulsion droplets \cite{Guttman:2016,Guttman:2017,Guttman:2019,Denkov:2015,Cholakova:2016,Denkov:2016,Cholakova:2019,Marin:2020}, as well as the theoretical models that us  \cite{GarciaAguilar:2021} and others \cite{Haas:2017,Haas:2019} have introduced to account for these observations. {\em 1)} because of the different surfactants used in Refs.~\cite{Guttman:2016,Guttman:2017,Guttman:2019} and \cite{Denkov:2015,Cholakova:2016,Denkov:2016,Cholakova:2019}, the physical mechanisms underpinning the shape transformations may in fact differ, in spite of the extraordinary resemblance in the experimental output. {\em 2)} the theoretical models introduced in Refs. \cite{GarciaAguilar:2021} and \cite{Haas:2017,Haas:2019} are mathematically equivalent, by virtue of the small magnitude of the stretching and gravitational energies considered in Ref. \cite{GarciaAguilar:2021}. In this Reply, we argue that neither of these hypotheses is well justified. 

\begin{figure}[t!]
\centering
\includegraphics[width =\columnwidth]{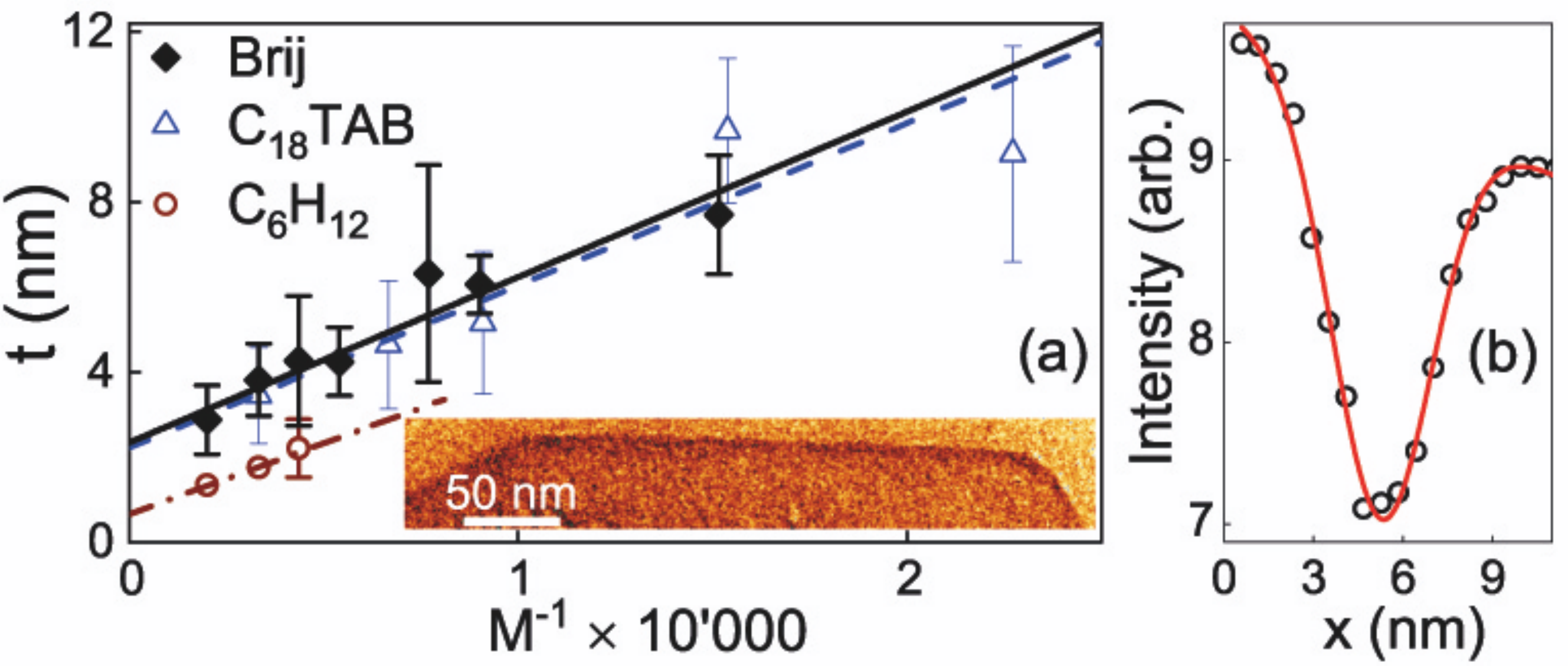}
\caption{\label{Fig1} (a) The cryoTEM interfacial widths $t$ of the faceted C$_{16}$ emulsion droplets, stabilized with C$_{18}$TAB \cite{Guttman:2019} or with Brij 78 surfactant (see legend), perfectly coincide. A much thinner interfacial width is obtained for a C$_6$H$_{12}$ emulsion, where no interfacial freezing takes place. Inset: The cryo-TEM image of the edge of a Brij - stabilized faceted droplet. The ``glow'' lookup table is employed. (b) CryoTEM intensity profile across an interface of a faceted Brij-stabilized droplet. The average fitted FWHM are shown in (a).} 
\end{figure}

To test the first hypothesis, we have performed direct cryoTEM imaging of Brij 78 non-ionic surfactant-stabilized emulsions, as those used in Refs. \cite{Denkov:2015,Cholakova:2016}. The interface of a faceted alkane [CH$_3$(CH$_2$)$_{14}$CH$_3$, denoted as C$_{16}$] droplet, suspended in a 1.5\% w/w aqueous Brij solution (inset to Fig.~\ref{Fig1}a), clearly demonstrates the absence of any surface-adjacent structure, such as the rotator crystals hypothesized in Refs.~\cite{Denkov:2015,Haas:2017}. The only \blue{detectable} feature  is a clearly defined interfacial layer, as that observed in C$_{18}$TAB-stabilized emulsions discussed in Ref. \cite{GarciaAguilar:2021}. To extract the interfacial layer's thickness, we fit the intensity profiles across the interfaces \blue{with a tilted} Gaussian function (Fig. ~\ref{Fig1}b). The full width at half maximum (FWHM) varies with the magnification $M$, averaging to $t=2.9 \pm 0.2$~nm at the highest accessible magnification. This $t$ value  matches the previously-estimated thickness of a monolayer~\cite{Guttman:2019}. Furthermore, as in our previous studies~\cite{Guttman:2019}, we linearly extrapolate the experimental FWHM values to $M^{-1}\rightarrow 0$ (Fig.~\ref{Fig1}a). The corresponding interfacial thickness of Brij-stabilized system, $2.3 \pm 0.2$~nm, perfectly agrees with the $2.2 \pm 0.9$~nm value previously reported for interfacially-frozen C$_{18}$TAB-stabilized C$_{16}$ emulsions~\cite{Guttman:2019}, indicating that only one crystalline monolayer is present at the interface of these faceted droplets, for either of the surfactants. Our analysis is further validated by the much smaller $t=0.7 \pm 0.1$~nm of the C$_{18}$TAB-stabilized cyclohexane (C$_6$H$_{12}$) emulsion, where no interfacial freezing takes place and the droplets are rounded (open circles in Fig.~\ref{Fig1}a). 

With respect to the second hypothesis, we notice that the numbers $\Delta\mathcal{E}_{S}$ and $\Delta\mathcal{E}_{G}$, used by Haas {\em et al}. to conclude that stretching and gravity are unimportant, correspond to energy minima and are not representative of the entire energy landscape, hence cannot be used to support the authors' conclusions. Specifically, the stretching energy of a spherical crystal depends on the configuration of the topological defects and ranges from the numbers $\mathcal{E}_{S}$ in Table I of Ref. \cite{GarciaAguilar:2021} to infinity (as any two of the twelve seed disclinations approach each other). Thus, stretching cannot be excluded {\em a priori} and its effect is, in fact, pivotal for the emergence of the icosahedral structure, where the twelve seed disclinations are maximally spaced. Similarly, because of the {\em quartic} dependence on $r=RH_{0}$, where $R$ is the droplet radius and $H_{0}$ the spontaneous curvature, the gravitational energy $\mathcal{E}_{G}$ is a rapidly increasing function of $R$. In the example considered by Haas {\em et al}. $R=10\,\mu$m \cite{Haas:2021}, which, using $H_{0}^{-1}\approx 60$~nm and $\Pi\approx 10^{-8}$~\cite{GarciaAguilar:2021}, gives $r \approx 167$ and $\Pi\Delta\mathcal{E}_{G}r^{4}\approx 27$. Thus {$\Pi\Delta\mathcal{E}_{G}r^{4}$ is much smaller than the dimensionless bending energy difference $\Delta\mathcal{E}_{H}r \approx 9 \times 10^{3}$ \cite{Haas:2021}. Yet, it is sufficient to take a droplet of radius $R=45\,\mu$m ($r=750$), thus well within the experimental range, i.e. $1-150\,\mu$m (Fig. 1f in Ref.~\cite{GarciaAguilar:2021}), for $\Pi\Delta\mathcal{E}_{G}r^{4}$ to have the same order of magnitude of $\Delta\mathcal{E}_{H}r$. For the largest droplets ($R=150\,\mu$m,  $r=2500$), $\Pi\Delta\mathcal{E}_{G}r^{4}$ is one order of magnitude {\em larger} than $\Delta\mathcal{E}_{H}r$, emphasizing the importance of gravity in the system.

\acknowledgements

We are grateful to Moshe Deutsch and Ray Goldstein for discussion. This work is partially supported by Netherlands Organisation for Scientific Research (NWO/OCW), as part of the D-ITP program (I.G.A. and L.G.), the Vidi scheme (P.F. and L.G.), the Frontiers of Nanoscience program (L.G.) and the Israel Science Foundation, grant no. 1779/17 (E.S.).

\end{document}